\documentclass[aps, prd, amsmath, floats, floatfix, twocolumn, nofootinbib, showpacs]{revtex4-1}
\usepackage{graphicx}
\usepackage{color}

\newcommand{\be}{\begin{eqnarray}}
\newcommand{\ee}{\end{eqnarray}}

\begin{document}

\title{Shadow of a dressed black hole and determination of spin and viewing angle}

\author{Lingyun Yang}

\author{Zilong Li}
\email[Corresponding author: ]{zilongli@fudan.edu.cn}

\affiliation{Center for Field Theory and Particle Physics and Department of Physics,
Fudan University, 220 Handan Road, 200433 Shanghai, China}

\date{\today}

\begin{abstract}
Shadows of black holes surrounded by an optically thin emitting medium have been extensively discussed in the literature. The Hioki-Maeda algorithm is a simple recipe to characterize the shape of these shadows and determine the parameters of the system. Here we extend their idea to the case of a dressed black hole, namely a black hole surrounded by a geometrically thin and optically thick accretion disk. While the boundary of the shadow of black holes surrounded by an optically thin emitting medium corresponds to the apparent photon capture sphere, that of dressed black holes corresponds to the apparent image of the innermost stable circular orbit. Even in this case, we can characterize the shape of the shadow and infer the black hole spin and viewing angle. The shape and the size of the shadow of a dressed black hole are strongly affected by the black hole spin and inclination angle. Despite that, it seems that we cannot extract any additional information from it. Here we study the possibility of testing the Kerr metric. Even with the full knowledge of the boundary of the shadow, those of Kerr and non-Kerr black holes are very similar and it is eventually very difficult to distinguish the two cases.
\end{abstract}


\maketitle

%%%%%%%%%%%%%%%%%%%%%%%%%%%%%%%

\section{Introduction}

The direct image of a black hole surrounded by an optically thin emitting medium is characterized by the presence of the so-called ``shadow'', namely a dark area over a brighter background~\cite{falcke}. The shape of the shadow corresponds to the apparent photon capture sphere as seen by a distant observer and it is the result of the strong light bending in the vicinity of the black hole. An accurate observation of the shadow can potentially provide information on the spacetime geometry around the compact object and test general relativity~\cite{sh1,sh1a,sh1b,aaa,sh1c}. The interest in this topic is today particularly motivated by the possibility of observing the shadow of SgrA$^*$, the supermassive black hole candidate at the center of the Milky Way, with sub-millimeter very long baseline interferometry (VLBI) facilities~\cite{doeleman}.

If the black hole has an optically thick and geometrically thin accretion disk, its apparent image is substantially different. However, even this ``dressed'' black hole is characterized by a shadow, whose shape is still determined by the spacetime geometry around the compact object and the viewing angle with respect to the line of sight of the distant observer~\cite{fukue,code}. In this case, the shape of the shadow corresponds to the apparent image of the inner boundary of the accretion disk, which, under certain conditions, should be located at the innermost stable circular orbit (ISCO)~\cite{steiner}.

In the Kerr metric, the shape of the shadow is only determined by the values of the black hole spin parameter $a_* = a/M = J/M^2$, where $M$ and $J$ are the black hole mass and spin angular momentum, and of the angle $i$ between the spin and the line of sight of the distant observer. The size of the shadow on the observer's sky is also regulated by the black hole mass and distance. In Ref.~\cite{maeda}, Hioki and Maeda proposed a simple algorithm to characterize the shape of the shadow of a Kerr black hole surrounded by an optically thin medium in terms of a distortion parameter $\delta$ and of the shadow radius $R$. If we independently know the mass and the distance of the object, the measurement of $\delta$ and $R$ can provide an estimate of the black hole spin and inclination angle. This algorithm has been later employed for the shadow of non-Kerr black holes as a general approach to characterize their shadow and get a rough estimate of the possibility of testing general relativity with future VLBI observations~\cite{sh2,naoki}.

With the same spirit, here we introduce two parameters, $\alpha$ and $\beta$, to characterize the shape of the shadow of a dressed black hole. Even in this case, we show that we can eventually measure the black hole spin and the inclination angle. However, it seems we cannot do more. We study the possibility of testing the Kerr metric by considering a non-Kerr background with a spin and a deformation parameter to quantify possible deviations from the Kerr solution. Even introducing a third parameter to describe the shadow shape, we fail to break the degeneracy between the spin and the deformation parameter of the background metric. We also introduce the function $R(\phi)$ to describe the entire shadow shape. Even with $R(\phi)$, we are not able to unambiguously distinguish a Kerr and a non-Kerr black hole. A similar problem was already found in the case of the shadow of a black hole surrounded by an optically thin emitting medium~\cite{naoki}, but in that case the shape and the size of the shadow do not change very much if the values of the physical parameters of the system vary. For that of a dressed black hole, both the shape and the size significantly change with the spacetime geometry and the inclination angle. Such a result can be probably understood by noting that the boundary of the shadow of a dressed black hole is the apparent image of the ISCO, whose radius is just one parameter. For a Kerr black hole, it is only determined by the spin of the object. For a non-Kerr black hole, it depends on both the spin and the deformation parameter and therefore there is a degeneracy between these two quantities, in the sense that the same ISCO radius can be obtained from different combinations of the spin and the deformation parameter.

We note that the results of our work have no implications for future observations of SgrA$^*$, as this object has no thin accretion disk. The shadow of a dressed black hole can be potentially observed in the case of stellar-mass black holes in X-ray binaries with X-ray interferometric techniques.
{The stellar-mass black hole candidate with the largest angular size on the sky should be that in Cygnus~X-1, which can be found in the state with a thin accretion disk and it seems thus a good candidate for this purpose.}
However, X-ray interferometric observations will be unlikely possible in the near future.

The content of the paper is as follows. In Section~\ref{s-2}, we briefly describe our approach to numerically compute the shadow of a dressed black hole. In Section~\ref{s-3}, we introduce the parameters $\alpha$ and $\beta$ to characterize the shape of the shadow and infer the black hole spin and its inclination angle with respect to the observer's line of sight. In Section~\ref{s-4}, we discuss how the detection of a shadow may be used to test the Kerr metric and we introduce the function $R(\phi)$ to describe its whole boundary. Summary and conclusions are reported in Section~\ref{s-5}. Throughout the paper, we employ units in which $G_{\rm N} = c = 1$.

\begin{figure*}
\begin{center}
\includegraphics[type=pdf,ext=.pdf,read=.pdf,width=8cm]{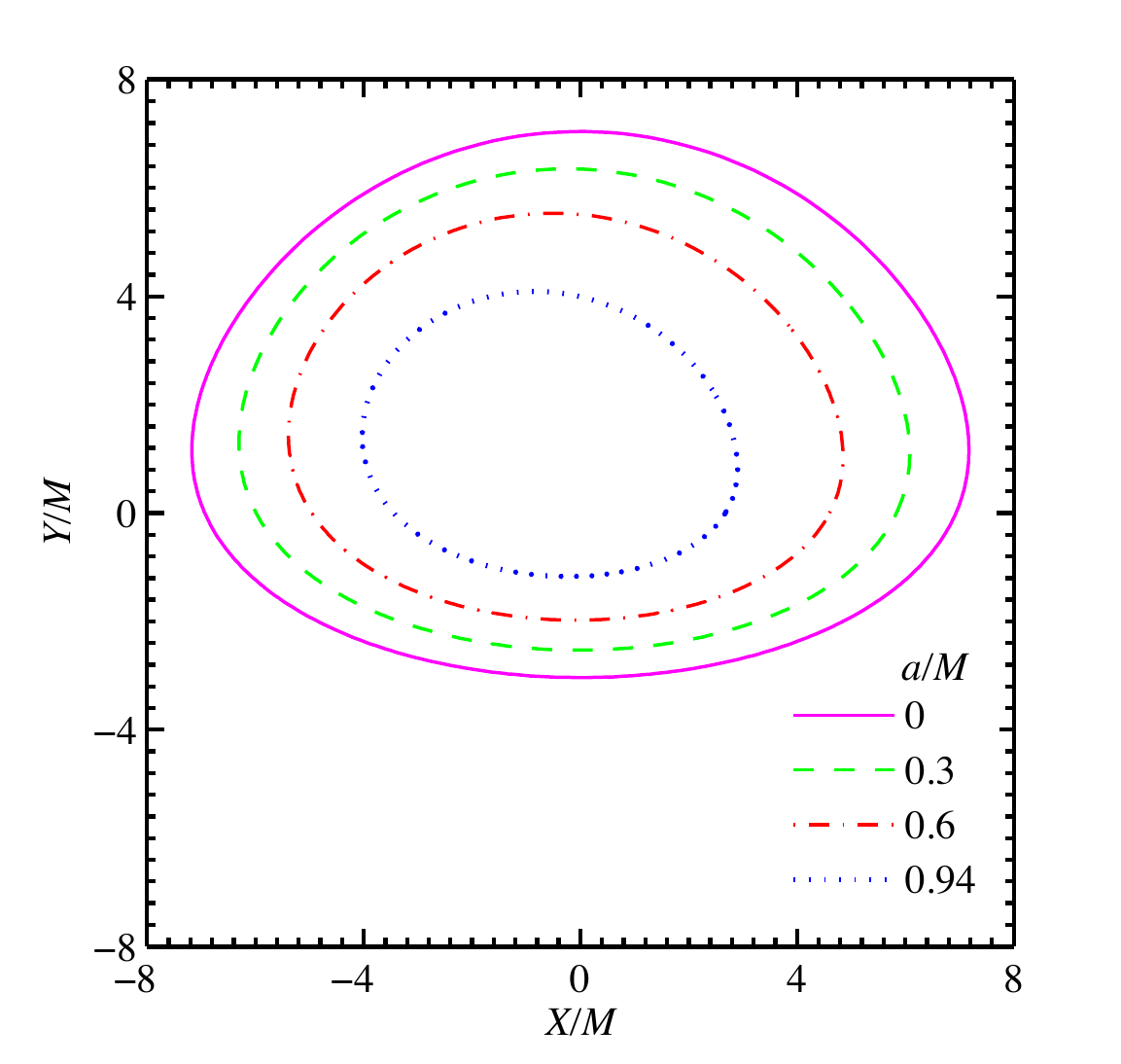}
\hspace{0.5cm}
\includegraphics[type=pdf,ext=.pdf,read=.pdf,width=8cm]{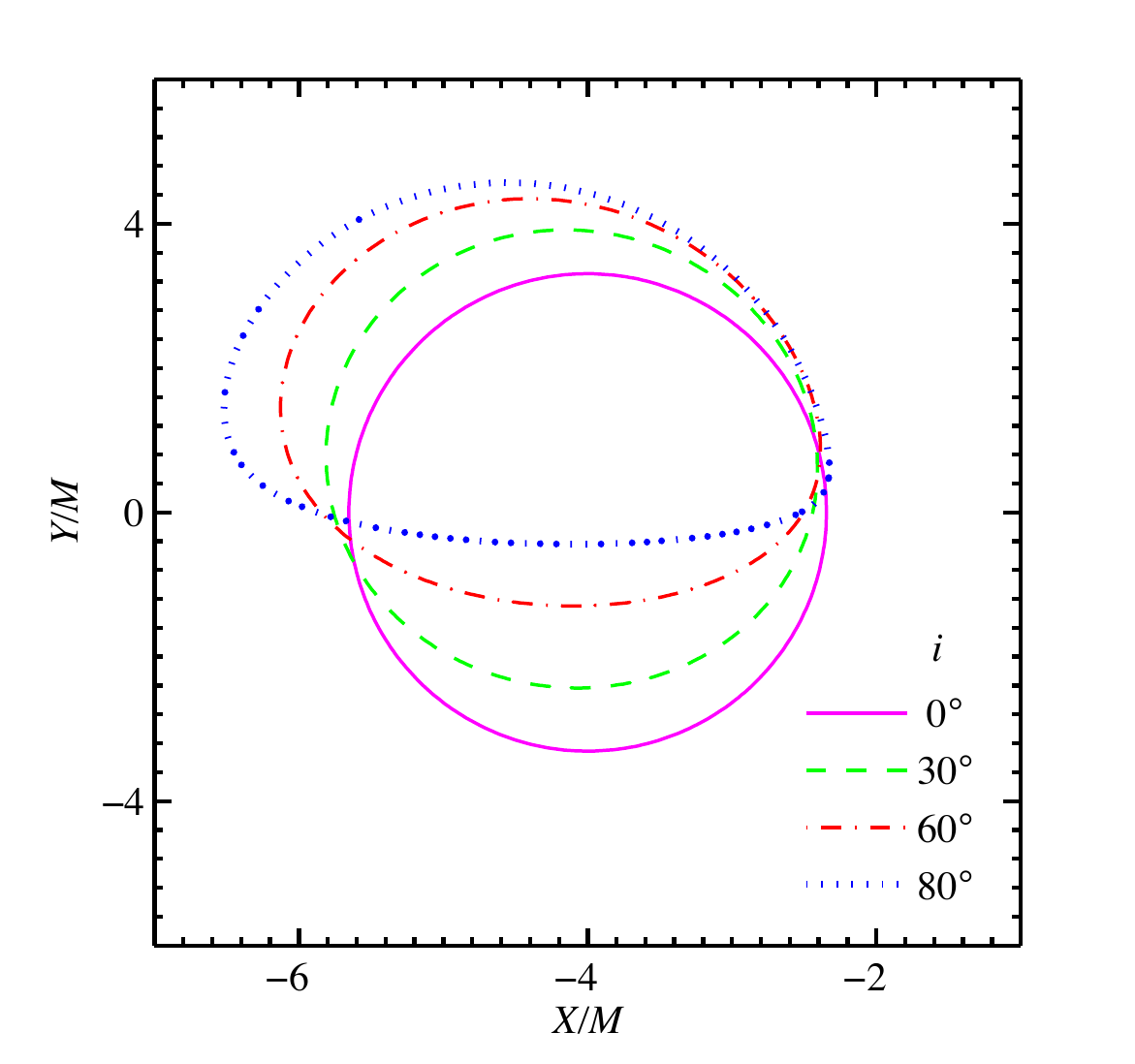}
\end{center}
\caption{Examples of shadows of dressed black holes. In the left panels, the inclination angle is $i=60^\circ$ and we show the effect of the spin parameter $a_*$. In the right panel, the spin parameter is $a_* = 0.9$ and we change the inclination angle $i$.}
\label{fig1}
\end{figure*}

\section{Calculation method \label{s-2}}

The Kerr solution is a Petrov type-D spacetime and therefore for a suitable choice of coordinates (e.g. in Boyer-Lindquist coordinates) the equations of motion are separable and of first order. This can somehow simplify the calculations of the black hole shadow. However, here we use the general approach and therefore its extension to non-Kerr backgrounds is straightforward. We use the code described in Ref.~\cite{code}. Since we are only interested in the calculation of the shape of the shadow, not in the intensity map of the direct image, we need to compute the photon trajectories from the image plane of the distant observer to the vicinity of the black hole and check whether the photon hits the thin accretion disk or not. The set of points on the observer's sky whose photons do not hit the disk forms the shadow of the black hole. Such photons can either hit the black hole or cross the equatorial plane between the inner edge of the disk and the black hole and then escape to infinity.

The initial conditions $(t_0, r_0, \theta_0, \phi_0)$ for the photon with Cartesian coordinates $(X,Y)$ on the image plane of the distant observer are~\cite{code}
\be
t_0 &=& 0 \, , \\
r_0 &=& \sqrt{X^2 + Y^2 + D^2} \, , \\
\theta_0 &=& \arccos \frac{Y \sin i + D \cos i}{\sqrt{X^2 + Y^2 + D^2}} \, , \\
\phi_0 &=& \arctan \frac{X}{D \sin i - Y \cos i} \, .
\ee
The initial 3-momentum of the photon, $\bf{k}_0$, is perpendicular to the plane of the image of the observer. The initial conditions for its 4-momentum are thus~\cite{code}
\be
k^r_0 &=& - \frac{D}{\sqrt{X^2 + Y^2 + D^2}} |\bf{k}_0| \, , \\
k^\theta_0 &=& \frac{\cos i - D \frac{Y \sin i + D
\cos i}{X^2 + Y^2 + D^2}}{\sqrt{X^2 + (D \sin i - Y \cos i)^2}} |\bf{k}_0| \, , \\
k^\phi_0 &=& \frac{X \sin i}{X^2 + (D \sin i - Y \cos i)^2} |\bf{k}_0| \, , \\
k^t_0 &=& \sqrt{\left(k^r_0\right)^2 + r^2_0  \left(k^\theta_0\right)^2
+ r_0^2 \sin^2\theta_0  (k^\phi_0)^2} \, .
\ee
In our calculations, the observer is located at $D = 10^6$~$M$, which is far enough to assume that the background geometry is flat. $k^t_0$ is thus obtained from the condition $g_{\mu\nu}k^\mu k^\nu = 0$ with the metric tensor of a flat spacetime. The photon trajectory is
 {calculated by solving the second order geodesic equations with the fourth order Runge-Kutta-Nystr\"om method, as described in~\cite{code}. The trajectory is}
numerically integrated backwards in time to check whether the photon hits the black hole, hits the disk, or crosses the equatorial plane between the inner edge of the disk and the black hole and then escapes to infinity. We note that some photons may cross the equatorial plane between the inner edge of the disk and the black hole and then hit either the disk or the black hole. We also note that we employ the usual set-up, in which the disk is on the equatorial plane and the inner edge of the disk is at the ISCO radius.

Fig.~\ref{fig1} shows some examples of our calculations. In the left panel, the inclination angle is fixed to $i = 60^\circ$ and we change the value of the spin parameter to see its effect on the shape of the shadow. The impact of $a_*$ is definitively different from the case of the shadow of a black hole surrounded by an optically thin emitting medium. In the case of a dressed black hole, the size of the shadow is very sensitive to the black hole spin, as it could have been expected from the fact the ISCO radius ranges from 6~$M$ for a non-rotating Schwarzschild black hole to $M$ for a maximally rotating Kerr black hole and a corotating disk, while the ISCO is at 9~$M$ when the disk of the maximally rotating Kerr black hole is counterrotating. The right panel in Fig.~\ref{fig1} shows the effect of the inclination angle. Here the black hole has spin parameter $a_* = 0.9$ and the inclination angle is $i = 0^\circ$, $30^\circ$, $60^\circ$, and $80^\circ$. Now the size of the shadow is roughly the same, while the shape changes significantly. If we look at the shadows of black holes surrounded by an optically thin emitting medium (see e.g. the figures in~\cite{maeda}), it is easy to conclude that both the shape and the size of the shadow of a geometrically thin and optically thick disk are much more sensitive to the values of the spin and of the inclination angle.

\begin{figure}[b]
\begin{center}
\includegraphics[type=pdf,ext=.pdf,read=.pdf,width=8cm]{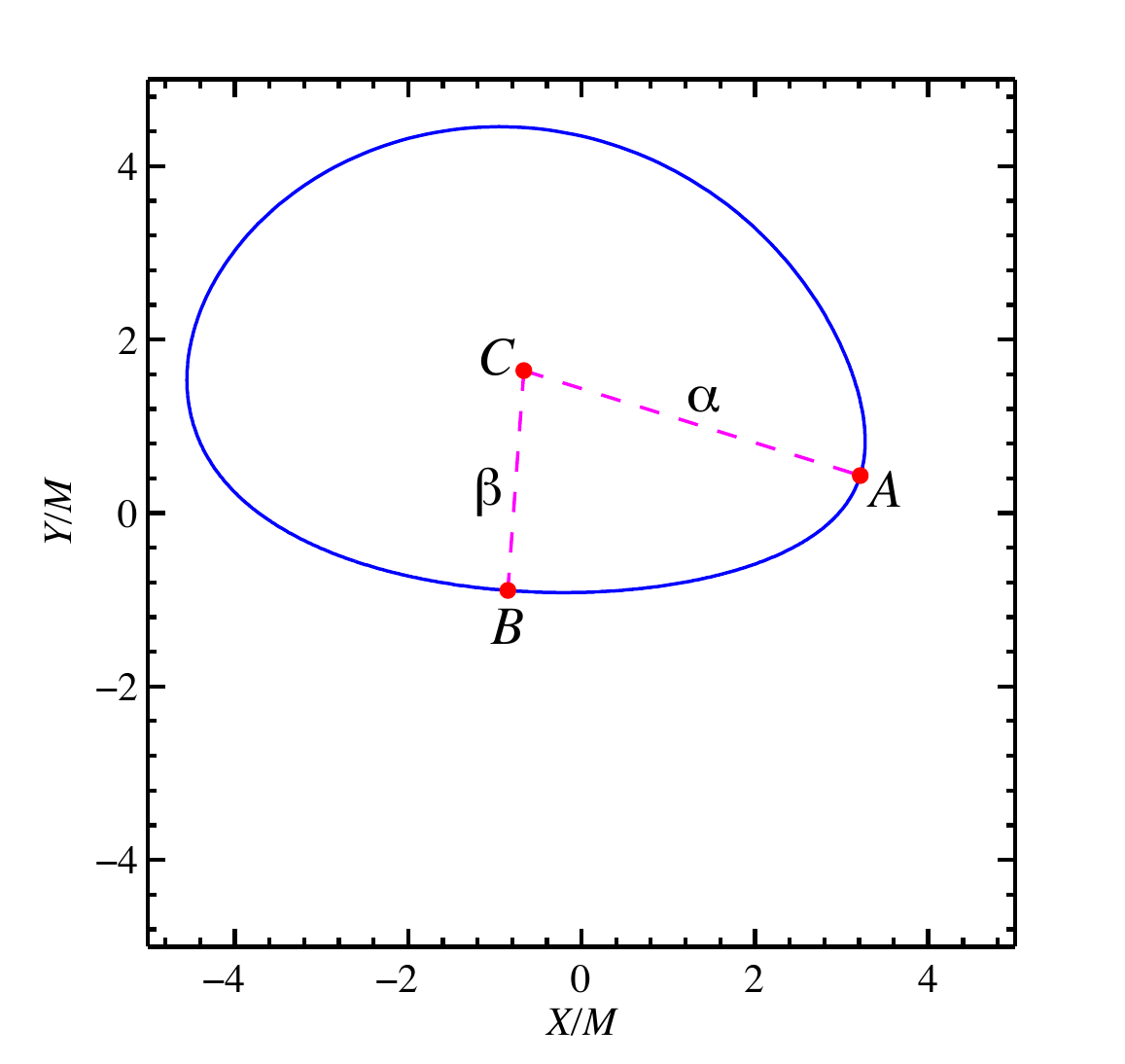}
\end{center}
\caption{Definition of the parameter $\alpha$ and $\beta$ to characterize the shape of the shadow of a dressed black hole.}
\label{fig2}
\end{figure}

\begin{figure*}[t]
\begin{center}
\includegraphics[type=pdf,ext=.pdf,read=.pdf,width=8cm]{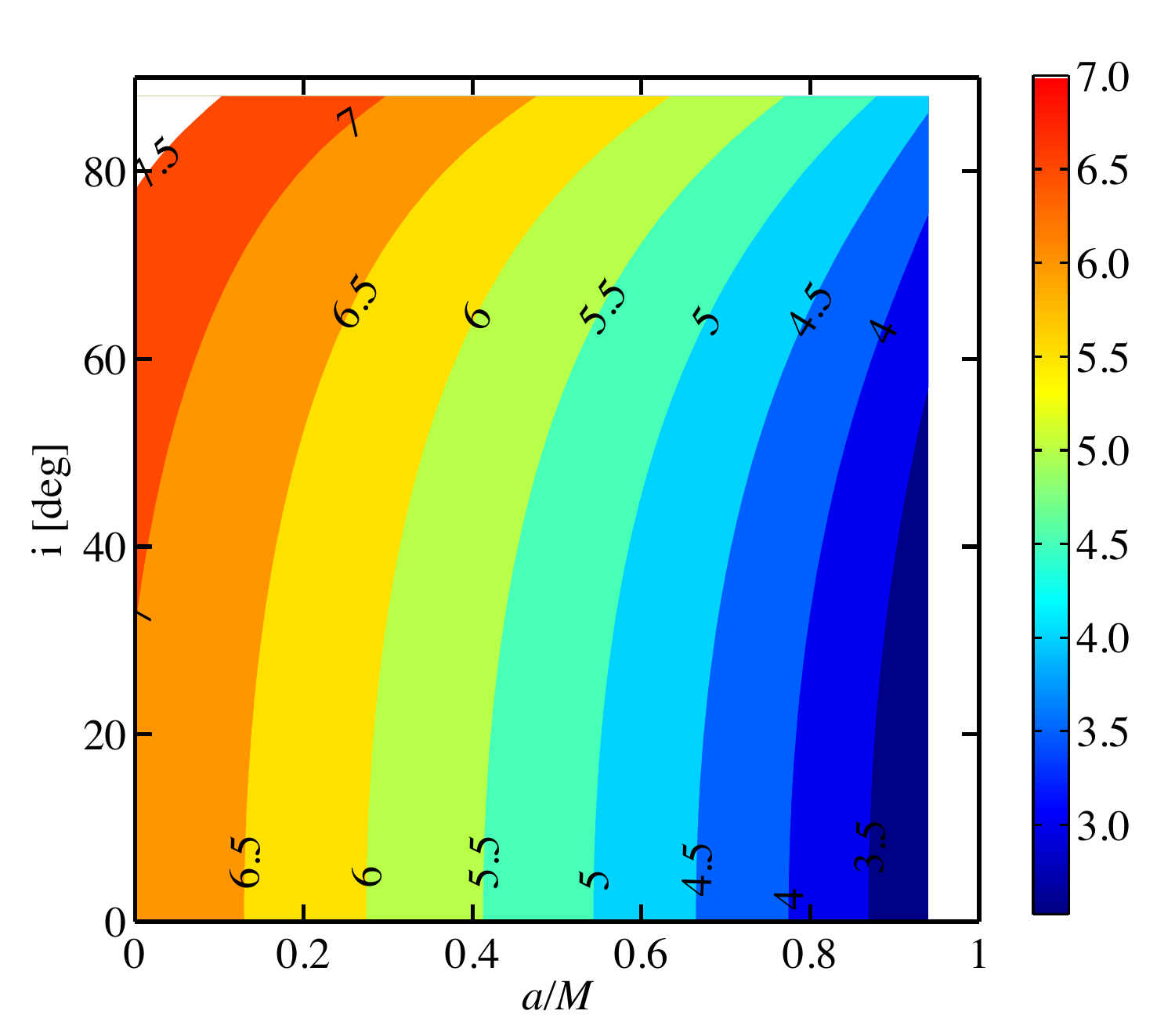}
\hspace{0.5cm}
\includegraphics[type=pdf,ext=.pdf,read=.pdf,width=8cm]{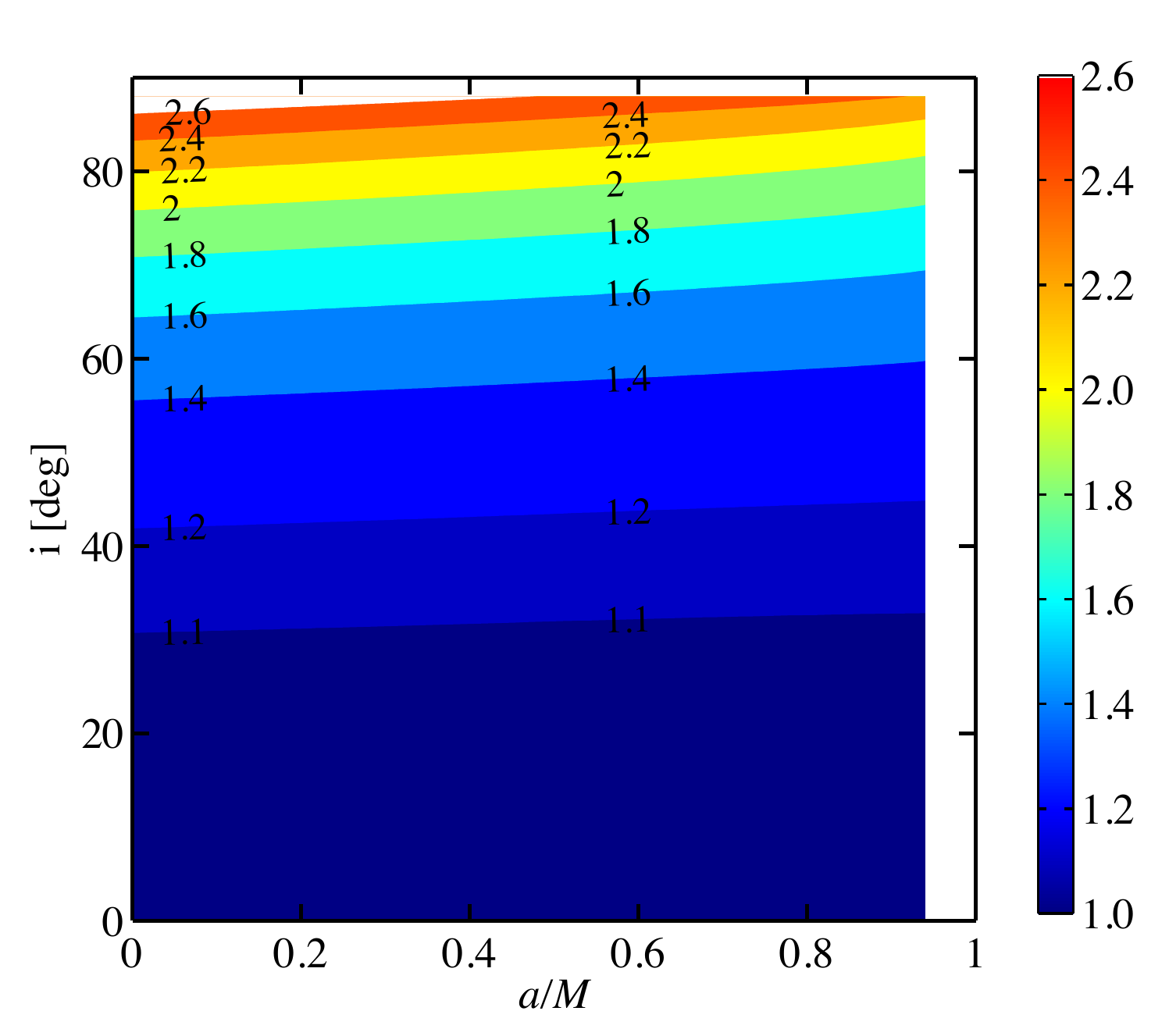}
\end{center}
\caption{Contour maps of $\alpha/M$ (left panel) and of $\alpha/\beta$ (right panel) in the plane spin parameter vs inclination angle for the Kerr metric. See the text for more details. }
\label{fig3}
%\end{figure*}
\vspace{0.5cm}
%\begin{figure*}
\begin{center}
\includegraphics[type=pdf,ext=.pdf,read=.pdf,width=8cm]{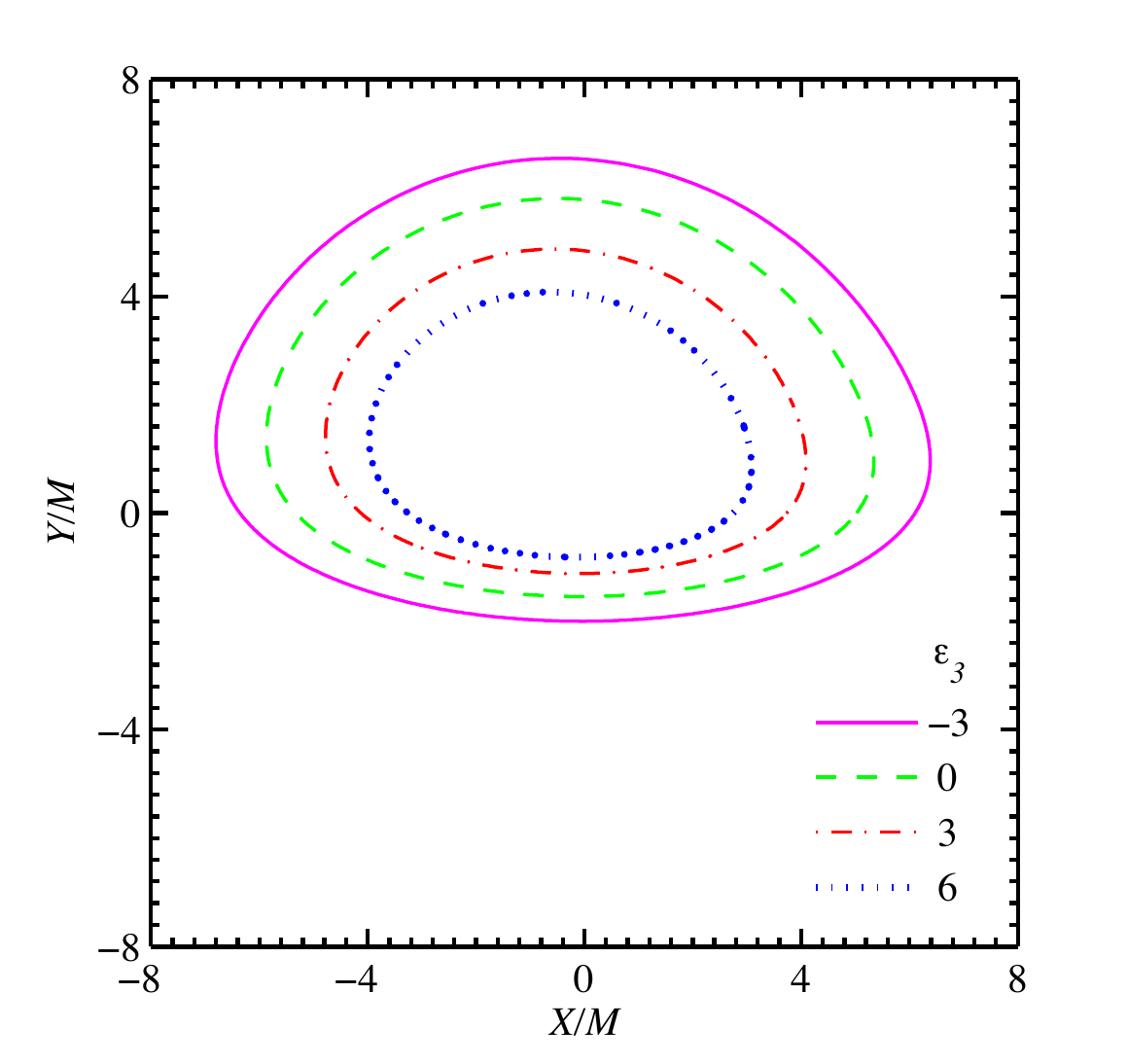}
\hspace{0.5cm}
\includegraphics[type=pdf,ext=.pdf,read=.pdf,width=8cm]{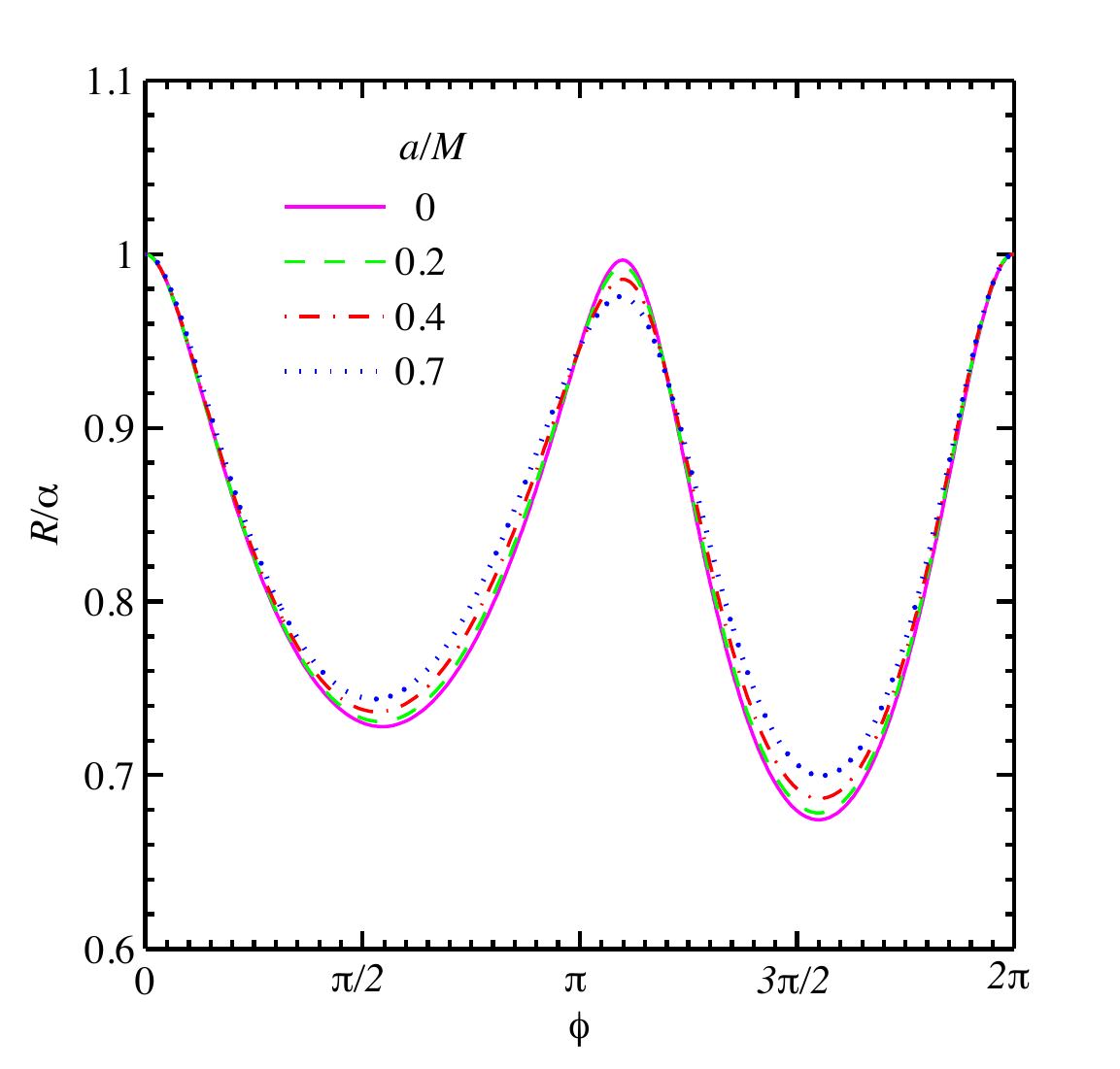}
\end{center}
\caption{Left panel: impact of the deformation parameter $\epsilon_3$ on the shape of the shadow of a dressed black hole. The inclination angle is $i = 70^\circ$, the spin parameter is $a_* = 0.5$, and we change the deformation parameter $\epsilon_3$. Right panel: examples of profile of the boundary of the shadow. Here we assume the Kerr metric, the inclination angle is $i=60^\circ$, and we show the profile for a few different spin parameters.}
\label{fig4}
\end{figure*}

\begin{figure}[t]
\begin{center}
\includegraphics[type=pdf,ext=.pdf,read=.pdf,width=8cm]{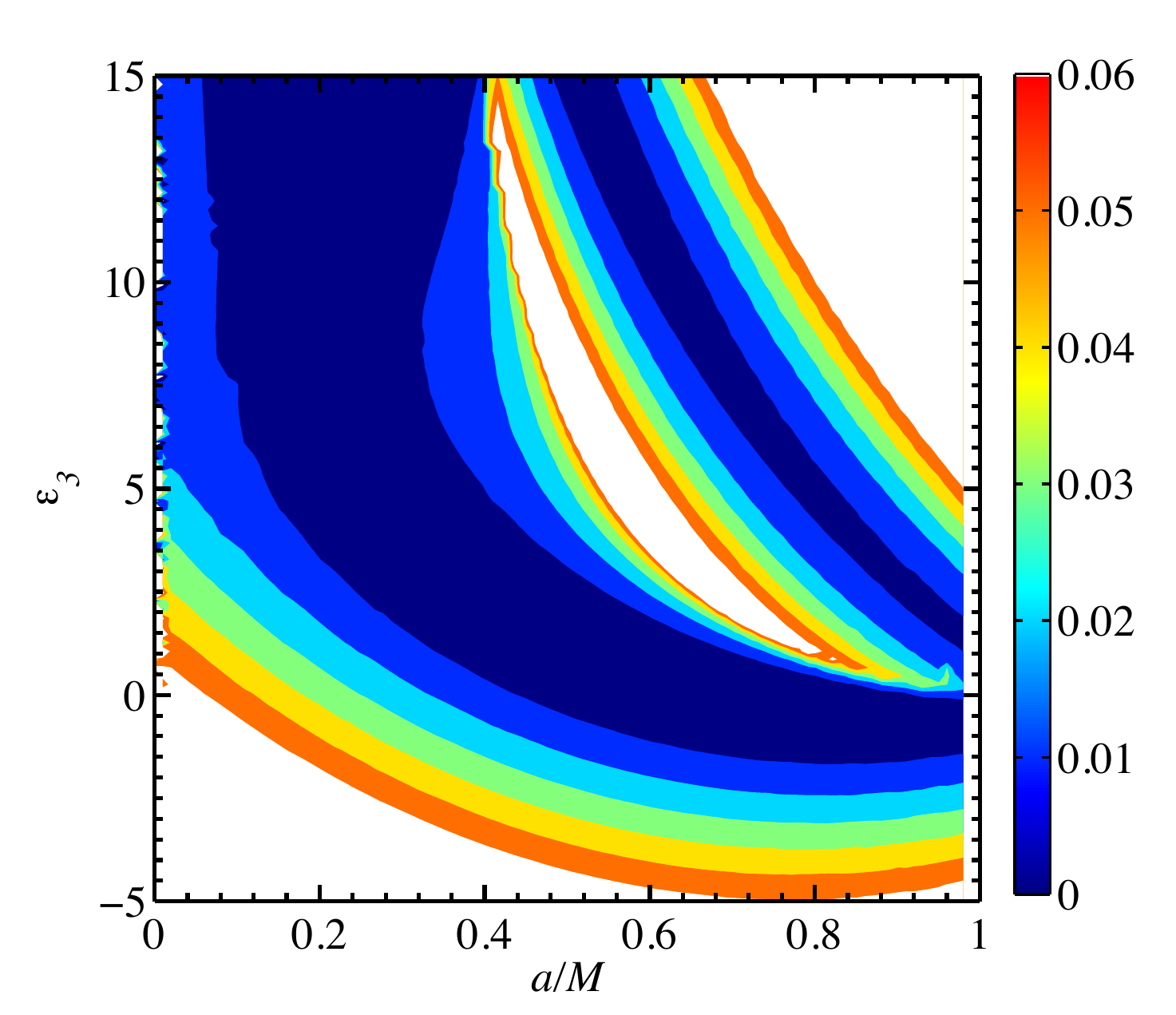}
\end{center}
\caption{Contour map of $S_1 (a_*, \epsilon_3, i)$ in which the reference model is a Kerr black hole with spin parameter 0.7 and observed from an inclination angle $60^\circ$. Here $i = 60^\circ$ is fixed. See the text for more details.}
\label{fig5}
\end{figure}

\begin{figure*}
\begin{center}
\includegraphics[type=pdf,ext=.pdf,read=.pdf,width=8cm]{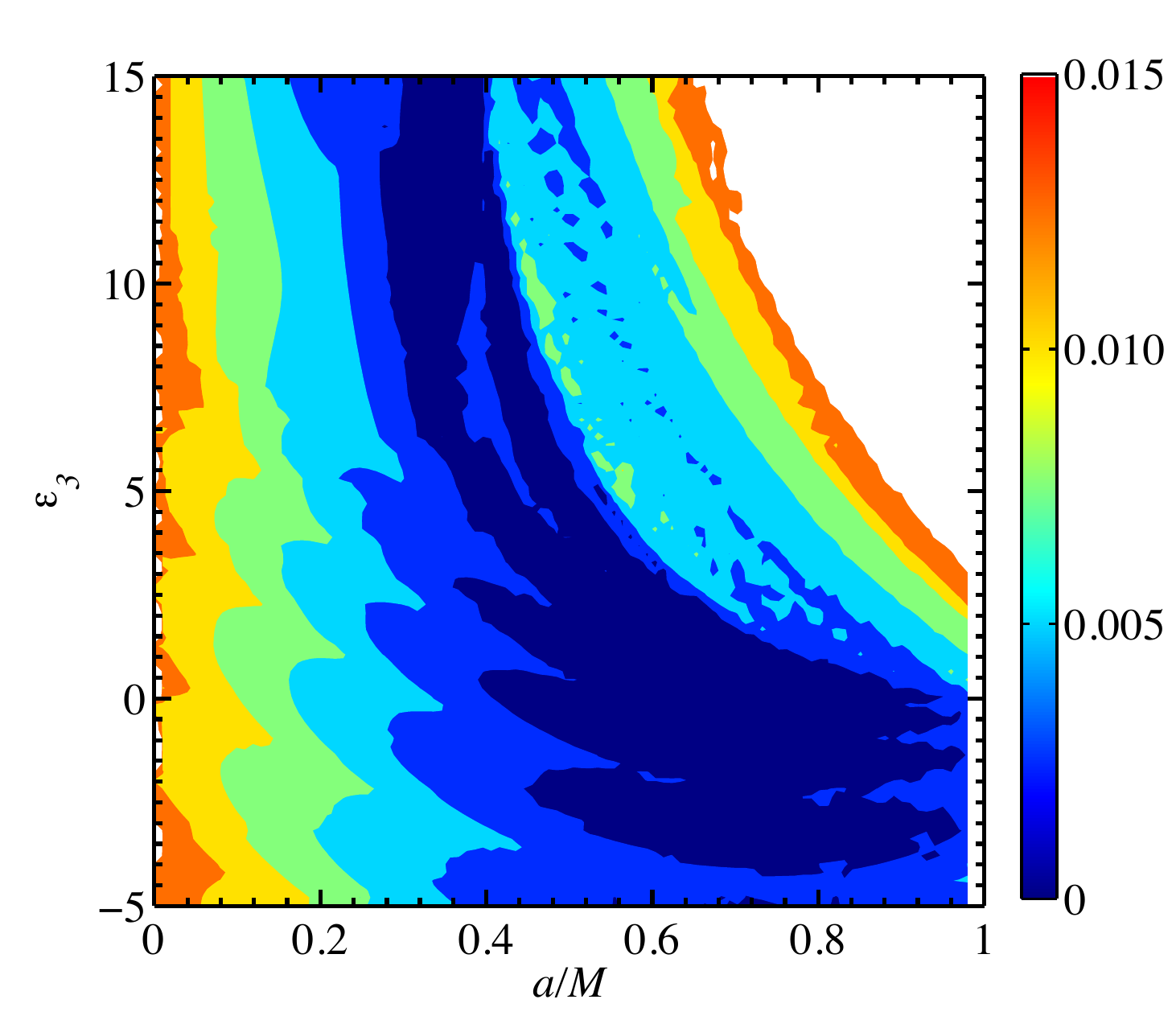}
\hspace{0.5cm}
\includegraphics[type=pdf,ext=.pdf,read=.pdf,width=8cm]{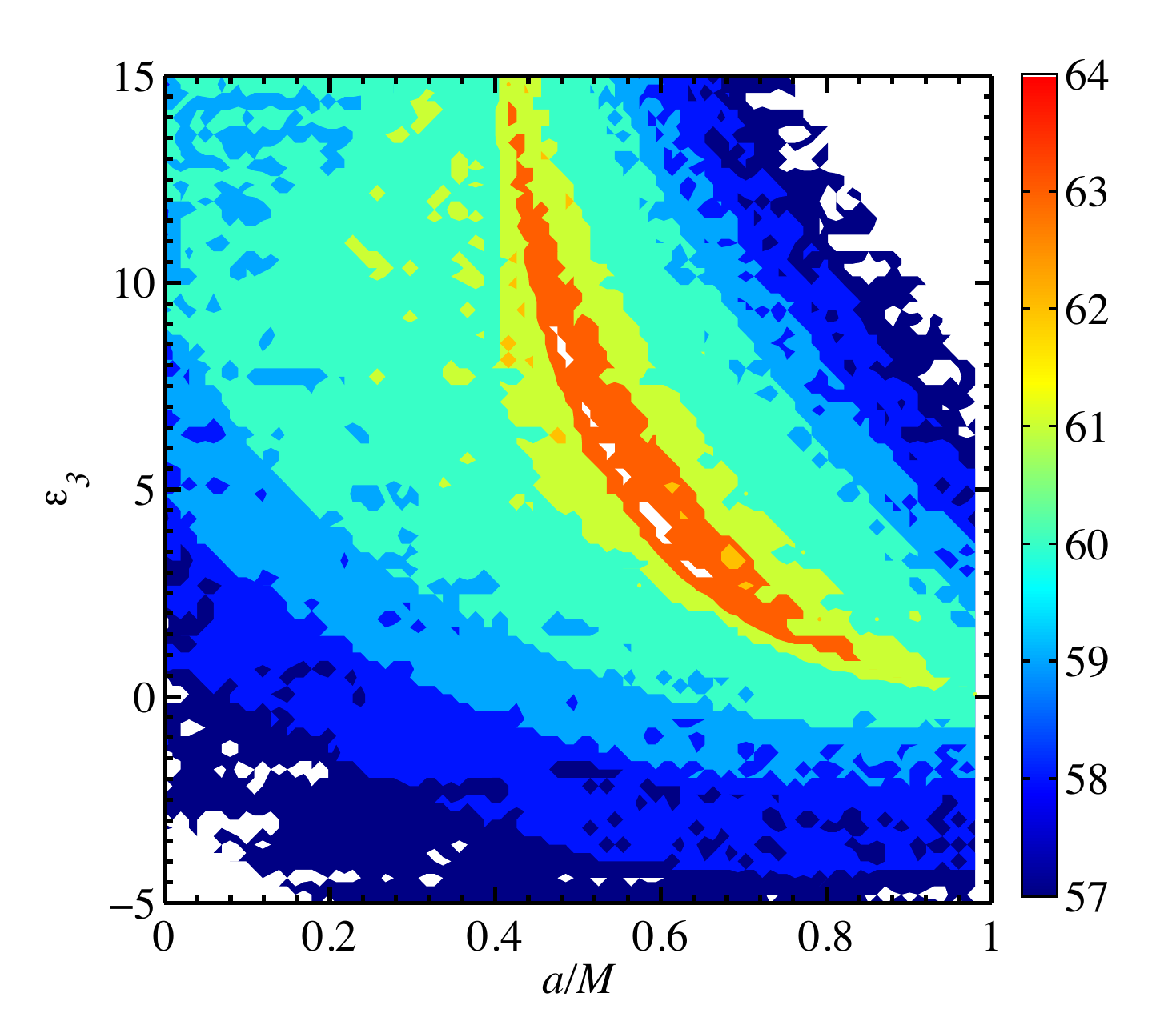}
\end{center}
\caption{Left panel: as in Fig.~\ref{fig5} with $i$ free in the fit. Right panel: contour map of the values of the inclination angle $i$ that minimizes $S_1$ in the left panel. See the text for more details.}
\label{fig6}
%\end{figure*}
\vspace{0.5cm}
%\begin{figure*}
\begin{center}
\includegraphics[type=pdf,ext=.pdf,read=.pdf,width=8cm]{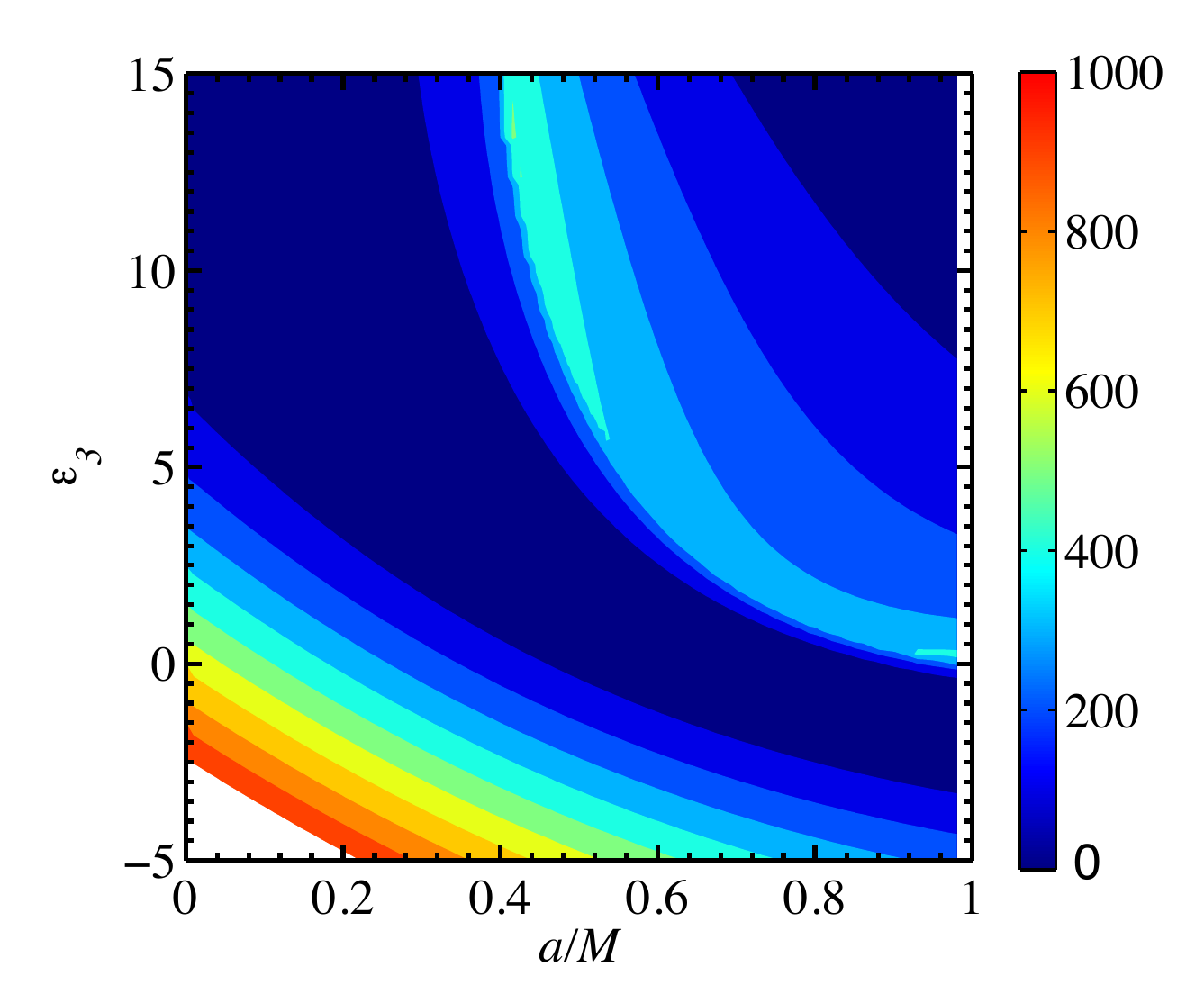}
\hspace{0.5cm}
\includegraphics[type=pdf,ext=.pdf,read=.pdf,width=8cm]{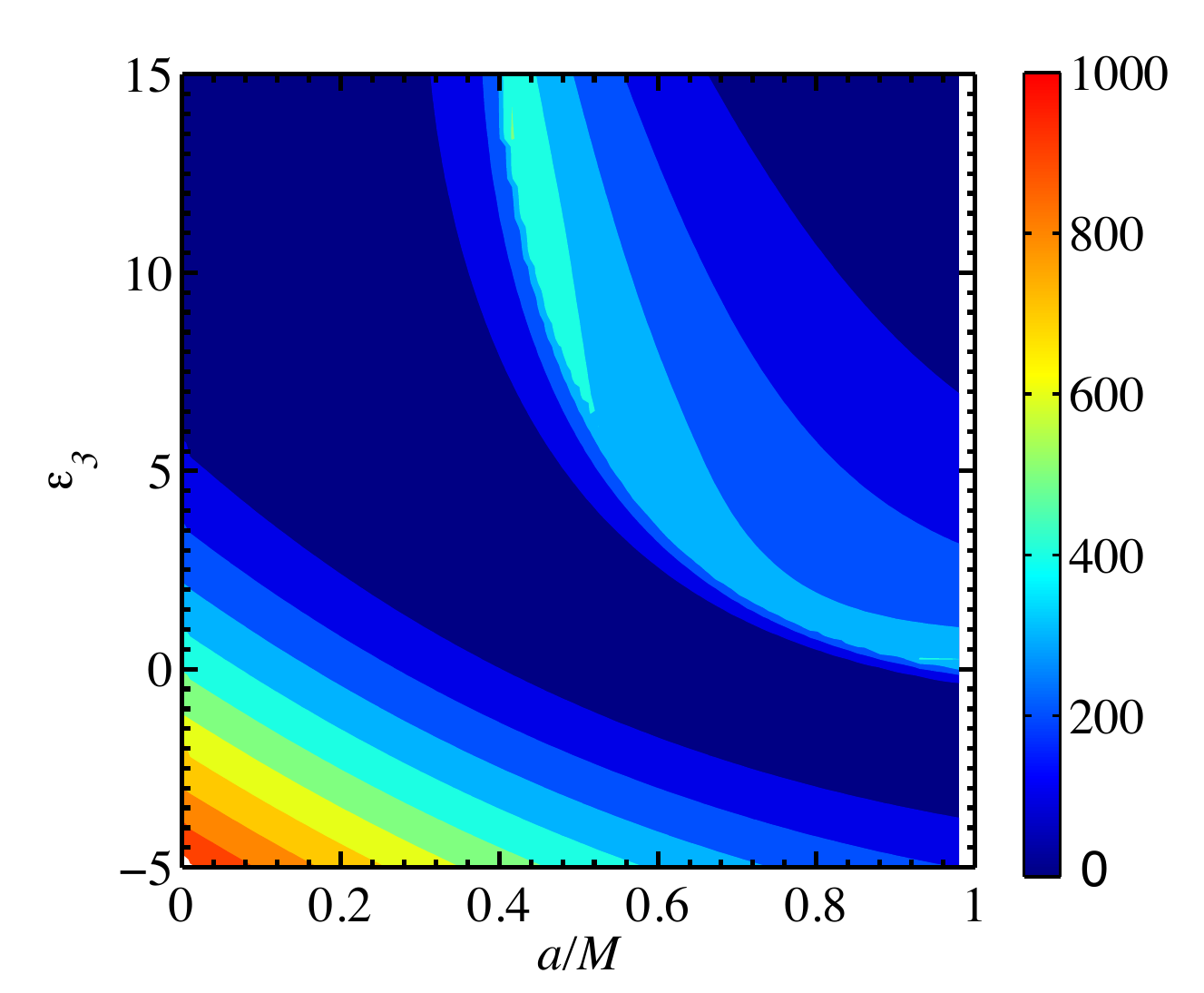}
\end{center}
\caption{Left panel: as in Fig.~\ref{fig5} for $S_2$. Right panel: as in the left panel in Fig.~\ref{fig6} for $S_2$. See the text for more details.}
\label{fig7}
\end{figure*}

\section{Determination of spin and viewing angle \label{s-3}}

In this section, we want to introduce two parameters to characterize the shadow of a dressed black hole, to be used to infer the values of its spin and viewing angle. Unlike the shadow of a black hole surrounded by an optically thin emitting medium, in general our shadows have not an axis of symmetry and therefore we need to adopt a slightly different approach with respect to that in Ref.~\cite{maeda}.

As first step in our algorithm, we find the ``center'' $C$ of the shadow (see Fig.~\ref{fig2}). It reminds the center of mass of a body and its coordinates $(X_C,Y_C)$ on the sky are given by
\be
X_C &=& \frac{\int \rho(X,Y) X dX dY}{\int \rho(X,Y) dX dY} \, , \nonumber\\
Y_C &=& \frac{\int \rho(X,Y) Y dX dY}{\int \rho(X,Y) dX dY} \, ,
\ee
where $\rho (X,Y) = 1$ inside the shadow and 0 outside. Once we have $C$, we can determine the distance of $C$ from every point of the boundary of the shadow. If $A$ and $B$ are the points on the boundary respectively with the maximum and minimum distance from $C$, we call $\alpha$ the distance $AC$ and $\beta$ the distance $BC$, as shown in Fig.~\ref{fig2}.

In the case of the Hioki-Maeda algorithm for the shadow of a black hole surrounded by an optically thin emitting medium, we determine two parameters: the shadow radius in units of the apparent size of the gravitational radius on the observer's sky, $R/M$, and the (dimensionless) distortion parameter, $\delta$. With these two quantities we can infer the black hole spin parameter, $a_*$, and the angle between the spin axis and the line of sight of the distant observer, $i$. Here we have the same situation. $\alpha/M$ is the counterpart of $R/M$, while $\alpha/\beta$ plays the role of the Hioki-Maeda distortion parameter $\delta$. In Fig.~\ref{fig3}, we show the contour maps of $\alpha/M$ (left panel) and $\alpha/\beta$ (right panel). $\alpha/M$ is mainly determined by the black hole spin, while the effect of the inclination angle $i$ is smaller. On the contrary, $\alpha/\beta$ is very sensitive to the exact value of $i$ and it is affected only weakly by $a_*$. If we can determine both $\alpha/M$ and $\alpha/\beta$, we can infer $a_*$ and $i$. We note that in our case of the shadow of a dressed black hole we could estimate the inclination angle $i$ (with some uncertainty) without knowing the apparent size of the gravitational radius on the observer's sky.

\section{Testing the Kerr metric \label{s-4}}

If we characterize the shape of the shadow of a black hole with two parameters, their determination can be used at most to infer two physical parameters of the black hole, like the spin and the inclination angle. In this section we want to figure out if the detection of the shadow of a dressed black hole can test the Kerr metric. To do this, we relax the assumption of the Kerr background and we consider a metric more general than the Kerr solution. The metric is now described by the black hole mass $M$, the spin parameter $a_*$, and, in the simplest case, a deformation parameter that quantifies possible deviations from the Kerr metric. We want to see if it is possible to measure the shadow and infer the three parameters of the system.

As example, we consider the Johannsen-Psaltis metric~\cite{jp-m}, whose line element reads
\be
\hspace{-0.5cm}
ds^2 &=& - \left(1 - \frac{2 M r}{\Sigma}\right) (1 + h) dt^2
+ \frac{\Sigma (1 + h)}{\Delta + a^2 h \sin^2\theta } dr^2 \nonumber\\
&& + \Sigma d\theta^2
- \frac{4 a M r \sin^2\theta}{\Sigma} (1 + h) dt d\phi \nonumber\\
&& + \Bigg[ \sin^2\theta \left(r^2 + a^2 + \frac{2 a^2 M r \sin^2\theta}{\Sigma} \right) \nonumber\\
&& \hspace{0.6cm} + \frac{a^2 (\Sigma + 2 M r) \sin^4\theta}{\Sigma} h \Bigg] d\phi^2 \, .
\ee
Here $\Sigma = r^2 + a^2 \cos^2\theta$, $\Delta = r^2 - 2 M r + a^2$, and, in the simplest version with only one deformation parameter, $h$ is
\be
h = \frac{\epsilon_3 M^3 r}{\Sigma^2} \, .
\ee
$\epsilon_3$ is the ``deformation parameter'' and it is used to quantify possible deviations from the Kerr geometry. The compact object is more prolate (oblate) than a Kerr black hole for $\epsilon_3 > 0$ ($\epsilon_3 < 0$); when $\epsilon_3 = 0$, we exactly recover the Kerr solution. The impact of the deformation parameter on the black hole shadow is shown in the left panel in Fig.~\ref{fig4}, where $a_*$ and $i$ are fixed and we change $\epsilon_3$.

First, we tried to identify a third parameter to characterize the shape of the shadow of a dressed black holes. We studied a few options (area of the shadow, distance between $C$ and other points on the boundary, etc.). However, we failed, in the sense that we did not find three parameters of the shadow to infer the three physical parameters of the black hole because of the degeneracy between $a_*$ and $\epsilon_3$.

To increase our chances of success, we map the whole shadow boundary. We introduce the function $R(\phi)$ defined as the distance between $C$ and the boundary of the shadow, starting from $\alpha$, i.e. $R(\phi=0) = \alpha$, where $\phi$ is the angle between the the segment $AC$ determining $\alpha$ and the segment between $C$ and the point under consideration. If we do not have an independent measurement of the black hole mass and distance, we can only measure the actual shape of the shadow (not the size) and therefore we can measure $R/\alpha$. Some examples are shown in the right panel in Fig.~\ref{fig4}, where we have only considered Kerr black holes with inclination angle $i=60^\circ$ and we vary the spin.

With the use of $R$, we can compare the shadow of different black holes. To quantify the similarity between two systems, we use the following estimator
\be
\hspace{-0.8cm}
S_1(a_*, \epsilon_3, i) = \sum_k \left(
\frac{R(a_*, \epsilon_3, i; \phi_k)}{\alpha(a_*, \epsilon_3, i)}
- \frac{R^{\rm ref}(\phi_k)}{\alpha^{\rm ref}}\right)^2 \, ,
\ee
where $R(a_*, \epsilon_3, i; \phi_k)$ is the function $R$ at $\phi = \phi_k$ of the black hole under consideration, $\alpha(a_*, \epsilon_3, i)$ is its $\alpha$, while $R^{\rm ref}(\phi_k)$ and $\alpha^{\rm ref}$ are, respectively, the the function $R$ at $\phi = \phi_k$ of some reference black hole and the corresponding $\alpha$.
 {With this estimator, we are simply considering the least squares method for the normalized radius of the shadow and therefore the ``best fits'' is obtained when the sum of the squared residuals is minimum. }

As example, we consider a reference black hole with $a_*=0.7$, $\epsilon_3 = 0$ (Kerr black hole), and $i=60^\circ$. Fig.~\ref{fig5} shows $S_1$ with $i=60^\circ$ for every black hole. In the left panel in Fig.~\ref{fig6}, we show $S_1$ with $i$ free and we have selected the inclination angle that minimizes $S_1$. In the right panel, we show the values of the inclination angles of the left panel. As we can see from Figs.~\ref{fig5} and \ref{fig6}, there is a strong correlation between the spin $a_*$ and the deformation $\epsilon_3$ and it seems it is very difficult to infer the actual values of the physical parameters of the system.

If we have an independent measurement of the black hole mass and distance, we can measure $R$ in units of gravitational radii. In this case, the estimator for the comparison of two shadows is
\be
\hspace{-0.8cm}
S_2(a_*, \epsilon_3, i) = \sum_k \frac{\left(
R(a_*, \epsilon_3, i; \phi_k)
- R^{\rm ref}(\phi_k)\right)^2}{M^2} \, .
\ee
 {Now the least squares method is for the absolute radius of the shadow}.
The results for constant $i$ and free $i$ are reported in Fig.~\ref{fig7}, respectively in left and right panels. Even in the case of the measurement of $R$, it seems we cannot test the Kerr metric because of the degeneracy between the spin and the deformation parameter. A simple explanation is probably that the shadow of a dressed black hole corresponds to the apparent image of the ISCO. The radius of the ISCO is just one parameter. If we have a Kerr black hole, it is only determined by the spin of the object and the measurement of the shadow can be used to infer the spin. In the case of a black hole with a spin and a possible non-vanishing deformation parameter, the same ISCO radius can be obtained with many different combinations of $a_*$ and $\epsilon_3$, namely there is a degeneracy. The measurement of the shape of the shadow cannot thus give an independent estimate of $a_*$ and $\epsilon_3$.

 {To test the Kerr metric, the shadow constrain should be combined with other observations that are not primarily sensitive to the position of ISCO in order to break the parameter degeneracy. While we have not investigated this point, we can expect that continuum-fitting and iron line measurements are not good to do it because they are strongly affected by the ISCO radius~\cite{cfm-iron}. Observations like measurements of QPOs~\cite{qpo} or estimate of the jet power~\cite{jet1} sounds more promising because based on other properties of the spacetime.}

\section{Concluding remarks \label{s-5}}

The shadow of a black hole is the dark area appearing in the direct image of the accretion flow. The shadows of black holes in general relativity and in alternative theories of gravity surrounded by an optically thin emitting medium have been extensively discussed in the literature and the interest on the topic is particularly motivated by the possibility of observing the shadow of SgrA$^*$ with VLBI facilities in the next few years. In this paper, we have studied the shadow of a dressed black hole, namely a black hole surrounded by a geometrically thin and optically thick accretion disk. Even these shadows can be potentially observed, but we will probably need to wait for a longer time because the sources are Galactic stellar-mass black holes in X-ray binaries, whose angular size on the sky is about five orders of magnitude smaller. X-ray interferometric techniques may observe these shadows, but there are no scheduled missions at the moment.

The boundary of the shadow of a dressed black hole corresponds to the apparent image of the inner edge of the disk, which, under certain conditions, should be located at the ISCO radius. Following the spirit of the Hioki-Maeda algorithm~\cite{maeda}, we have introduced the parameters $\alpha$ and $\beta$ to characterize the shape of the shadow. In the standard set-up with a Kerr black hole, the boundary of the shadow only depends on the black hole spin and inclination angle with respect to the line of sight of the distant observer, like the shadow of a black hole surrounded by an optically thin emitting medium. However, unlike the latter case, both the shape and the size significantly change if we vary $a_*$ and $i$. This may suggest that the shadow of a dressed black hole is more informative than that of a black hole surrounded by an optically thin emitting medium, but this is not what we have eventually found.

If the mass and the distant of the black holes are known, the measurement of $\alpha/M$ and $\alpha/\beta$ can be used to infer the black hole spin parameter $a_*$ and the inclination angle $i$. Actually the two estimates are very weakly correlated, because $\alpha/M$ is mainly sensitive to $a_*$ while $\alpha/\beta$ is essentially determined by $i$. As a result, if the mass and the distant of the black holes are not known and one can only measure $\alpha/\beta$, it is still possible to get an estimate of the inclination angle $i$ with some uncertainty.

As second step, we have checked if an accurate determination of the shadow of a dressed black hole can be used to test the Kerr metric. In the simplest case, the system is now defined by three physical parameters (spin parameter, deformation parameter, viewing angle). While it is an easy job to infer $a_*$ and $i$ in the standard set-up, it seems we cannot test the Kerr metric because of a degeneracy between the spin and the deformation parameter. Even the full knowledge of the boundary of the shadow, which we have here described with the function $R(\phi)$, cannot do it. In other words, the shadow of a dressed black hole is very sensitive to two parameters, the spacetime geometry around the compact object and the inclination angle, but not more.

%%%%%%%%%%%%%%%%%%%%%%%%%%%%%%%

\begin{acknowledgments}
We thank Cosimo Bambi and Jiachen Jiang for useful discussions and suggestions. This work was supported by the NSFC grant No.~11305038, the Shanghai Municipal Education Commission grant for Innovative Programs No.~14ZZ001, the Thousand Young Talents Program, and Fudan University.
\end{acknowledgments}

%%%%%%%%%%%%%%%%%%%%%%%%%%%%%%

\end{document}